\magnification=1200
\pretolerance=10000  
\def\folio{\ifnum\pageno=1\nopagenumbers\else\number\pageno\fi}
%
%

\def\lax    {\ifmmode{_<\atop^{\sim}}\else{${_<\atop^{\sim}}$}\fi}
\def\gax    {\ifmmode{_>\atop^{\sim}}\else{${_>\atop^{\sim}}$}\fi}
\def\kms    {\ifmmode{{\rm ~km~s}^{-1}}\else{~km~s$^{-1}$}\fi}

\def\lo     {~${\rm L}_{\odot}$}
\def\mo     {~${\rm M}_{\odot}$}
\def\moyr   {\hbox{~${\rm M}_{\odot}\,{\rm yr}^{-1}$}}
\def\etal   {{et~al.~}}

\def\bk{\lower 6pt\hbox{${\buildrel k\over \sim}$}}
\def\bv{\lower 6pt\hbox{${\buildrel v\over \sim}$}}

\def\blankline  {\vskip10truept}

\def\Sgr   {Sgr A$^\star~$}
\def\sgr   {Sgr A$^\star~$}

%
%
\def\singlespace {\smallskipamount=3pt plus1pt minus1pt
                  \medskipamount=6pt plus2pt minus2pt
                  \bigskipamount=12pt plus4pt minus4pt
                  \normalbaselineskip=12pt plus4pt minus4pt
                  \normallineskip=1pt
                  \normallineskiplimit=0pt
                  \jot=3pt
                  {\def\smallskip {\vskip\smallskipamount}}
                  {\def\medskip   {\vskip\medskipamount}}
                  {\def\bigskip   {\vskip\bigskipamount}}
                  {\setbox\strutbox=\hbox{\vrule
                    height8.5pt depth3.5pt width 0pt}}
                  \parskip 0pt
                  \normalbaselines}
%
%
\def\doublespace {\smallskipamount=6pt plus2pt minus2pt
                  \medskipamount=12pt plus4pt minus4pt
                  \bigskipamount=24pt plus8pt minus8pt
                  \normalbaselineskip=24pt plus8pt minus8pt
                  \normallineskip=2pt
                  \normallineskiplimit=0pt
                  \jot=6pt
                  {\def\smallskip {\vskip\smallskipamount}}
                  {\def\medskip   {\vskip\medskipamount}}
                  {\def\bigskip   {\vskip\bigskipamount}}
                  {\setbox\strutbox=\hbox{\vrule
                    height17.0pt depth7.0pt width 0pt}}
                  \parskip 12.0pt
                  \normalbaselines}
\hsize 6.1 truein
\vsize 9 truein
\hoffset 0 truein
\voffset 0 truein
%
\parindent 24.0pt

\blankline
\centerline {\bf COLLIDING WINDS IN THE STELLAR CORE}
\centerline {\bf AT THE GALACTIC CENTER:}
\centerline {\bf SOME IMPLICATIONS}
\vskip 0.3truein
\centerline{\bf Leonid M. Ozernoy$^{1,2}$, Reinhard Genzel$^3$,
and Vladimir V. Usov$^4$}
\vskip 0.2truein
\item {$^1$} Computational Sciences Institute and Department of Physics
\& Astronomy, George Mason U., Fairfax, VA 22030-4444, USA
\item {$^2$} Laboratory for Astronomy and Solar Physics, Goddard Space Flight
Center, Greenbelt, MD 20771, USA;
{\rm e-mail}: ozernoy@hubble.gmu.edu,
ozernoy@stars.gsfc.nasa.gov
\item {$^3$} Max-Planck-Institut f\"ur Extraterrestrische Physik,
Postfach 1603, Garching bei M\"unchen,
85740 Germany; {\rm e-mail}: genzel@mpe.mpe-garching.mpg.de
\item {$^4$} Department of Physics, Weizmann Institute, Rehovot 76100,
Israel; {\rm e-mail}: fnusov@weizmann.weizmann.ac.il

\blankline
\noindent{\bf ABSTRACT}

We point out that a high number density of stars in the core of 
a dense star cluster such as the central stellar
cluster at the Galactic center, where many stars possess strong stellar winds,
should result in collisions of those winds.
The wind collisions in the dense stellar core at the Galactic center would 
result in production of
strong X-ray flares with the rate of $\sim 10^{-4}~(N_w
/10^3)^2$ yr$^{-1}$
and duration of $\sim 1$ week, where $N_w$ is the number of the wind
producing stars in the core.
Presence of a massive black hole would enhance the
stellar density around it and would make collisions of the winds in the core
substantially more frequent.
Collisions of the stellar winds in the cluster
have also a number of interesting observable implications,
including generation of $\gamma$-rays by particles accelerated by the shocks
from the colliding winds. These processes are also expected to be relevant
to compact regions of intense star formation elsewhere. 
\blankline
\par\vfill\eject
\blankline
\noindent{\bf 1. INTRODUCTION}
\blankline
The first near-IR observations of a broad HeI emission line from extended
source IRS~16 at the Galactic center have indicated an intense
outflow of matter from it (Hall \etal\ 1982; Geballe \etal\ 1982).
Subsequent detection, in addition to the above HeI line, of broad Br$\alpha$
and Br$\gamma$ HI emission lines, basically confirmed that interpretation
(Geballe \etal\ 1984).
Detailed \hbox{near-IR} observations of the
Galactic center indicate that the broad line region as a source of mass
outflow  has the following approximate parameters:
$${\rm characteristic ~linear~ size~ (FWHM)}\,~ l \approx 3^{\prime\prime}
\approx 0.1~{\rm pc},$$
$${\rm gas~ density}~ n\approx 2\times 10^4~{\rm cm}^{-3},~~{\rm wind~
velocity}~v\approx700~{\rm  km~ s}^{-1},    \eqno (1)$$
$${\rm  mass-loss~ rate}~ \dot M\approx 4\times10^{-3}\hbox{\moyr}$$
(Geballe \etal\ 1987, 1991; Allen et al. 1990). If the wind consist mostly 
of helium, this can diminish $n$ by a factor of two.

This outflow  has been shown to arise from the sum of individual winds 
produced by hot stars of various locations and spectral types.
They contain $\sim$20 luminous blue supergiants  (Krabbe et al. 1991),
including several components of the central IRS~16 complex.
Their P Cygni profiles directly indicate the existence of strong supersonic
outflows (Najarro et al. 1994, Krabbe et al. 1995).
The temperatures and spectral types of the
above mentioned stars range from somewhat cooler, $\sim 20,000$ K blue
supergiants with strong $\lambda$2.058 $\mu$m He I line emission
to $\gax 30,000$ K WN- and WC-type Wolf-Rayet (WR) stars. Mass loss rates
of the individual stars range from a few $10^{-5}$ up to a few $10^{-4}$\moyr
at velocities of 300 to 900 km s$^{-1}$ (Najarro et al. 1993,
Krabbe et al. 1995, Blum et al. 1995, Tamblyn et al. 1996).

The presence of such powerful winds at the Galactic center has several
profound implications for the central activity as well as for many
morphological peculiarities found in the region:

\item {}$\bullet$ interception of part of the winds by a putative central
black hole (\Sgr) and the resulting wind accretion can serve as a
source of the black hole  feeding and accretion luminosity (Ozernoy 1989,
1993; Melia 1992; Narayan et al. 1995);

\item {}$\bullet$ interaction of the wind with the ionized and molecular gas
at the Galactic center can explain much of its morphology
(Yusef-Zadeh \& Wardle 1993);

\item {}$\bullet$ both momentum and energy of the wind create a
`wind-dominated' regime within the central few tenths of a parsec, which
influences substantially the dynamics of the region (Lutz et al. 1993).

\noindent
These and other wind features are reviewed, in more detail and both
from observational and  theoretical sides, by Genzel, Hollenbach, \&
Townes (1994). Recently we have shown that
the kinetic pressure of the wind is able to stop the inflow of matter
into the central parsec from the circumnuclear ring and to reduce drastically
the accretion rate onto the central black hole (Ozernoy \& Genzel 1996, 
1997). In the present paper, we would like to point out one more important
corollary of the extended wind at the Galactic center --{\it collisions
between the individual stellar winds within the core of the central stellar
cluster}. The existence of a very dense core has been recently demonstrated
by a high resolution imaging of the nuclear star cluster (Eckart et al. 1993).
Besides, both acceleration of particles by shock waves
from the colliding winds in the cluster and hard X-ray and
$\gamma$-ray emission of these particles are considered below.

\blankline
\noindent {\bf 2. COLLIDING WINDS IN THE STELLAR CORE}
\blankline
\noindent {\bf 2.1. The physics of the collision and the production of
X-rays}
\blankline
As is well known, among massive stars of WR and OB types, which have been
found to be strong X-ray emitters, the most powerful are those belonging
to binary systems.
This phenomenon, commonly interpreted as due to the collisions of the winds
from the components entering the binary, was at first predicted theoretically
(Prilutskii \& Usov 1976). These authors have shown that collision of the
wind  away from a WR star either with the wind from a companion OB star
or with the surface of the latter induces a shock that heats the gas up to
the temperature $T\sim 10^7$~K, resulting in X-ray emission with
luminosity $L_X\sim 10^{33}-10^{35}$ erg s$^{-1}$. By the present time,
the theory of this process has been elaborated in much detail and
it successfully serves to both interpret and guide the observations
(for reviews, see e.g. Pollock 1987, 1995; Usov 1992, 1995;
Stevens 1995).

It should be noted that the solitary massive OB stars also have inherent X-ray
emission whose origin is related to shocks that appear as a result
of instabilities in the
radiation-driven winds from massive stars (Owocki \& Rybicki 1985). However,
OB stars belonging to binary systems are usually several times brighter
at X-rays, and this has been explained by colliding
winds (Chlebowski 1989). Moreover, the binary's $L_X$ can be much
higher if the companion of the OB star is a WR star
rather than another OB star. This is due to the fact that the mass loss
rate and the terminal velocity of the matter outflow far from the star
are higher for WR stars
$$\dot M_{\rm WR}\sim (0.8-8)\times 10^{-5}~{\rm M_\odot~ yr}^{-1},~~
V_{\rm WR}^\infty \sim (1-5)\times 10^{3}~{\rm km~ s}^{-1} \eqno (2a)$$
than for OB stars of same masses
$$\dot M_{\rm OB}\sim  10^{-6}~{\rm M_\odot~ yr}^{-1},~~
V_{\rm OB}^\infty \sim (1-3)\times 10^{3}~{\rm km~ s}^{-1} \eqno (2b)$$
(e.g. Abbott \& Conti 1987, Leitherer 1988).

In the case when both the WR and OB winds collide
with their (supersonic) terminal velocities and X-ray absorption might
be neglected, the resulting X-ray emission
as a function of the binary's parameters is given by (Usov 1992):
$$L_X\simeq 10^{35}\left(\dot M_{-5}^{\rm WR}\right)^{1/2}
\left(\dot M_{-6}^{\rm OB}\right)^{3/2}
\left(V_{3}^{\rm WR}\right)^{1/2}
\left(V_{3}^{\rm OB}\right)^{-3/2}D_{13}^{-1}~\,{{\rm erg\, s}}^{-1}\,,
\eqno (3)$$
where $\dot M^{\rm WR}$ and $\dot M^{\rm OB}$ are normalized to $10^{-5}$
\moyr and $10^{-6}$\moyr, correspondingly; $V^{\rm WR}$ and $V^{\rm OB}$
are normalized to $10^3$ km s$^{-1}$, and the distance $D$ between WR and OB
stars is normalized to $10^{13}$ cm. The radiation-driven coronal winds from
massive stars attain their terminal velocity $V^\infty$ at the radius
$r_{\rm ter}$ which is a few
stellar photosphere radii (e.g. Friend \& MacGregor 1984).

The analysis of the conditions under which the stellar winds collide in
a WR + OB binary (Usov 1992) shows that neither a very close nor a
very wide binary is able to produce strong X-ray emission. This is because
in a very close binary the X-ray emission is strongly absorbed in the WR
and OB winds (Luo, McCray \& Mac Low 1990; Usov 1990; Stevens, Blondin \&
Pollock 1992), whereas in a very wide binary the X-ray luminosity
decreases with an increase of the distance $D$ between the components of the
binary: $L_X\propto D^{-1}$, or even faster ($L_X\propto D^{-4/3}$) when no 
equalization of the ion and electron temperatures occurs within at least one 
shock layer around the contact surface (Usov 1992). It is expected that 
the X-ray luminosity of a WR + OB binary reaches its maximum,
$L_X^{\rm max}\simeq 10^{35}$ erg s$^{-1}$, when $D\simeq\tilde D\equiv
(\dot M_{\rm WR}V_{\rm WR}^\infty/\dot M_{\rm OB}V_{\rm OB}^\infty )^{1/2}
r_{\rm ter}$. For a typical WR + OB binary one gets
$\tilde D\simeq 10^{13}$ cm within a factor of 2-3 or so (Usov 1992).

So far massive binaries have been observed neither in IRS 16 nor elsewhere
within the central parsec of the Galactic center, although lunar occultation
observation do not exclude that some of the IRS 16 components might be
multiple stars (Simons et al. 1990).
Meanwhile all known massive stars within the central parsec
are separated by such large distances ($\sim 10^{16} - 10^{17}$ cm) that the
effects of their {\it individual} wind-wind collisions are rather weak
(in contrast to the {\it collective} effects of wind collisions on the
surrounding ISM -- see Sec. 4). However,
those effects could be much stronger,
compared to IRS 16, in the central dense stellar cluster -- the nuclear
core. An inferred  very high total stellar density in
the core $\sim 10^7 -10^8$ stars/pc$^3$ (Eckart et al. 1993)
implies comparatively frequent stellar encounters, with a part of them 
being as close as $\sim 10^{13}$ cm to provide an effective collision of 
stellar winds and, as a result, the production
of strong X-ray emission, acceleration of particles, and generation of
related hard radiation (see below).
An interesting circumstance is that we do not need to know an (unknown)
number of WR + OB binaries: these close `pairs' are formed temporarily
during encounters in the cluster core to reveal this event by
producing the above phenomena and then the stars fly apart not to meet each
other anymore!

\blankline
\noindent {\bf 2.2. The rate of near-by stellar encounters
in the stellar core}
\blankline
To calculate the rate of wind collisions within the stellar core,
we introduce the number of wind producing stars, $N_w$. We assume that
the radial distribution of these, most massive and luminous, stars follows
an isothermal-like stellar distribution with a
density run found by Eckart et al. (1993):
   $$ n(r)=n_c (r_c/r)^2, \eqno (4) $$
where $r_c=0.15\pm 0.05$ pc is the core radius, $\rho_c=9\sigma_v^2/4\pi
Gr_c^2=
10^{7.9\pm 0.5}~{\rm M_\odot pc}^{-3}$ is the inferred mass density within
the core, and $\sigma_v=100\pm 25$ km s$^{-1}$ is the line-of-sight central
velocity
dispersion. In fact, available data indicate (Eckart et al. 1993, Genzel
et al. 1996) that the early type stars may have even a slightly smaller core
radius (0.1 pc) whereas for the late type stars it is larger ($r_c\sim
0.3-0.5$) pc.

While calculating the rate with which the winds in the core collide, we
constrain ourselves only with those encounters of the wind producing stars
whose collision periastron is $D\lax \tilde D \simeq 10^{13}$ cm.
The rate of stellar encounters with an impact parameter $D$ occuring in an
isothermal spherical cluster is given by
$${\cal R}\simeq 20 \pi^{3/2}n_w^2r_c^3\left[4D^2\sigma_v + {G(m_{w1}+m_{w2})
D\over\sigma_v}\right], \eqno (5)$$
where the second term in the brackets accounts for the gravitational focusing
of the encountering stars whose masses are $m_{w1}$ and $m_{w2}$.

In order to evaluate a characteristic value of ${\cal R}$, let us take
$m_{w1}=m_{w2}=10$\mo~ and put $n_w\equiv 10^7~n_7$ stars/pc$^3$ and
$r_c\equiv 0.1~ r_{-1}$ pc. One gets
$${\cal R}\simeq 3.8\cdot 10^{-10}n_7^2r_{-1}^3\left[4D^2_{13}\sigma_2+2.7
{D_{13}\over \sigma_2}\right]~{\rm s}^{-1}\sim 0.1~n_7^2r_{-1}^3~{\rm yr}
^{-1},   \eqno (6)$$
where $\sigma_2 \equiv \sigma_v/100~{\rm km~s^{-1}}$.
Given the number of stars producing winds $N_w$, their density  within the
core is
$n_w=N_w/{4\over 3}\pi r_c^3\simeq 2.4\cdot 10^5~N_3 r_{-1}^{-3}~
{\rm pc}^{-3}$, where $N_3\equiv N_w/10^3$. Eq. (6) yields then
$${\cal R}\sim 0.6\cdot 10^{-4}~N_3^2r_{-1}^{-3}~{\rm yr}^{-1}.
\eqno (7)$$

\blankline
\noindent {\bf 2.3. Parameters of X-ray flares due to colliding winds}
\blankline
To evaluate an actual collision rate, one needs to know the number of
wind producing stars within the core. Evidently, early-type
stars would be of prime interest here. Let us discuss available constraints
to the number of those stars.

The entire bolometric luminosity within $r\leq 0.6$ pc at the
Galactic center is $L_{\rm bol}\simeq 2\times 10^7$\lo~, with about 10\%
of it in the form of Ly$_{\rm c}$ photons (Zylka et al. 1995, Krabbe et al.
1995). The luminosity as high as this could be provided by just $\sim 10^2$
luminous stars like the Allen-Forrest (AF) object (Allen \etal 1990, Forrest
\etal 1987)
with $L_{\rm AF}\simeq (2-3)\times 10^{5}$\lo~
and mass-loss rate $\dot M\simeq (6\pm 1)\times 10^{-5}$\moyr~ (Najarro et al.
1994). However the stars as luminous as the AF star are rather rare.

Let us suppose, for a moment, that the `wind-dominating' region at the
Galactic center is created by much more numerous, although less luminous
but still massive, stars.
Using as a scaling an empirical relationship $\dot M\propto L^{1.25}$
(van Buren 1985) and normalizing it with the help of the AF star, one can
see that both the total mass-loss rate $\dot M\simeq 4\times 10^{-3}$\moyr~
[Eq. (1)] and the total luminosity
of the wind producing region could be reproduced by $N_w$ stars of
luminosity $L_w$ and mass-loss rate $\dot M_w$ each. The values of $N_w,
L_w,~ {\rm and}~ \dot M_w$ are presented in Table 1:                        

\centerline{\% PUT TABLE 1 HERE \%}

As Eq. (7) shows, even if the number of massive, wind producing stars in the
core were as large as $N_w=10^4$, the rate of near, $D\lax 10^{13}$ cm,
stellar encounters would only be
${\cal R}\sim 0.6\times 10^{-2}~{\rm yr}^{-1}$. Besides, those winds
would have $\dot M\simeq 2\times 10^{-7}$\moyr~, at which according to Eq.
(3) their collisional $L_X\ll 10^{35}$ erg/s. Only a combination of $N_w
\lax 10^3$ and $r_c\ll 0.1$ pc would make the dissipation of kinetic
energy of the colliding winds observable as X-ray flares with both
$L_X\sim 10^{35}$ erg s$^{-1}$ and high enough rate of these flares.
The duration of the flares is
$$\Delta t\simeq {D\over \sqrt 3 \sigma_v}\sim 1~{D_{13}\over \sigma_2}~
{\rm week}. \eqno (8)$$

Taking into account that $L_X\propto D^{-1}$ (see Eq. (3)), Eqs
(6) and (8) yield ${\cal R}\propto (\Delta L_X)^{-2}$ and
$\Delta t\propto (\Delta L_X)^{-1}$, where $\Delta L_X$ is
the amplitude of modulations of the X-ray luminosity of the compact cluster
at the Galactic center because of the wind collisions. For $N_w=10^3$, we have
numerically ${\cal R}\sim 10^{-4}(\Delta L_X/10^{35}\,{\rm erg\,s}^{-1})^{-2}
r^{-3}_{-1}$ yr$^{-1}$ and $\Delta t\sim 1 (\Delta L_X/10^{35}\,{\rm
erg\,s}^{-1})^{-1}\sigma_2^{-1}$ week. From both the condition ${\cal R}\simeq
(\Delta t)^{-1}$ and the fact that the soft X-ray luminosity of the Calactic
Center is $\sim 1.5\times 10^{35}$ erg s$^{-1}$, the X-ray flux from
the Galactic Center has to be variable on the time scale of one year with
the amplitide of $\sim 1$ \%.

Available high resolution near-infrared imaging data obtained with the MPE
camera SHARP (Eckart \etal 1993, 1995) enable us,
from counting the number of objects on the SHARP map,  to derive an upper
limit for the number of stars
earlier than B0.5/B1 [$K\sim 15,~ L\sim (5-10)\times 10^3$\lo].
Taking into account the uncertainty due to confusion, noise, and
completeness at $K\sim 15$, there are no more than $\sim$100 to 200 such
stars within $r\lax 4''$ of the central stellar core.
Having compared this upper limit with Eq. (7) one concludes
that detections of individual X-ray flares from collisions of stellar
winds in the dense core are untenable at the Galactic center.

The rate of wind collisions could be substantially increased if one
accounts for the influence of a putative massive black hole thought to
present at the Galactic center as a compact radio source \Sgr. The rate
of stellar collisions in a density spike around \Sgr is enhanced by a
factor of 30-40 (Ozernoy 1994). Even if the density spike has already
been destroyed by those frequent stellar collisions the gravitational
field of the black hole enhances stellar density of unbound stars in the
black hole vicinity and hence increases appreciably the rate of wind
collisions. Therefore, X-ray flares from the colliding winds at the
Galactic center, if detected, would provide an indirect evidence in favor
a massive black hole sitting there.

Unless the site for colliding stellar winds is more compact and/or
a massive black hole enhances the collision rate in its vicinity, a larger 
number of wind producing stars would result in an observable effect. Yet
the disentangling of the thermal, wind-produced X-ray flares from the
persistent X-ray emission turns out to be rather difficult partly because
that emission is due to the (shocked) global wind itself. Meanwhile
the above consideration, exemplified by the Galactic center case, reveals 
conditions under which the thermal X-ray flares produced by colliding winds
could be observable in some compact star clusters elsewhere. In the 
Galactic center, the overall soft X-ray emission due to the shocked winds
in the central cavity ($R\approx 1.5$ pc) is given by $L\approx 10^{35}\left
(p_{th}/10^8~{\rm cm^{-3}K}\right)^2f_V~{\rm erg~s^{-1}}$ (Genzel et al. 
1994), where $p_{th}\approx 10^8~{\rm cm^{-3}K}$ is the characteristic 
thermal pressure in the hot 
($T\simeq 10^7~{\rm K}$) shocked wind and $f_V$ is the volume filling factor 
of the shocked gas. Therefore, the persistent emission would be temporally
modulated by individual X-ray flares if the (otherwise similar) star cluster
had a smaller $p_{th}$ implying more rarefied gas.  

\blankline 
\noindent{\bf 3. NON-THERMAL EMISSION OF PARTICLES}

\noindent{\bf ACCELERATED IN SHOCKS FROM COLLIDING WINDS}
\blankline 

As is known, colliding winds which flow away from massive stars are able to
accelerate charged particles via shock waves. Indeed, several WR + OB
binaries demonstrate non-thermal radio emission that has been shown to be
associated with the accelerated particles
(Eichler \& Usov 1993 and references therein).  Spectral properties
of that radio emission are consistent with what would be expected from the
electrons which are accelerated by the shocks produced by colliding winds
(Eichler and Usov, 1993; Dougherty et al. 1996).

\bigskip
\noindent{\bf 3.1.Conditions for particle acceleration} 
\noindent {\bf and the outflowing gas structure}
\smallskip

To accelerate particles by shocks, the strength of the magnetic field
at the region of stellar wind collision, at the distance of a few
$10^{16}$ cm from a WR star, has to be within the range $10^{-12}\ll B \ll
3\times 10^{-3}$ Gauss [see eqs. (8)-(12) in Eichler and Usov 1993]. The
strength of the magnetic field in the plasma flowing away from a WR
star is given, far enough from the star ($r>RV^\infty_{WR}
/V_{\rm rot}$), by
$$B\simeq B_{_S}{V_{\rm rot}\over V^\infty_{WR}}{R^2\over r_{_A}r}\,
\eqno (9)$$
(e.g., Usov and Melrose 1992), where $B_{_S}$ is the surface magnetic field,
$R$ is the stellar radius, $V_{\rm rot}$ is the surface rotational velocity,
$r_{_A}$ is the Alfven radius, and $V^\infty_{WR}$
is the terminal velocity of the WR wind. For the feducial
values of $B_{_S}\sim 10^2$ G, $R\sim 10^{12}$ cm,
and $V_{\rm rot}\sim 0.1 V^\infty_{WR}$ at
$r\sim$ a few $\times 10^{16}$ cm, one gets $B\sim$ a few $\times
10^{-4}$~G, which is well within the range of $B$ given above. This and
the other conditions which are necessary for particle acceleration by shocks
(e.g., Eichler \& Usov, 1993) are met in the colliding winds
at the stellar core at the Galactic Center. Therefore, charged
particles (electrons, protons, and nuclei) can be accelerated in the
core to high enough energies. The maximum Lorentz factor of accelerated
electrons is about $10^5$ [see eqs. (13)-(14) in Eichler and Usov (1993)].

The colliding winds outflowing from the massive stars in the central
stellar cluster produce an ensemble of shock waves. Indeed, although both
the structure and kinematics of the interstellar matter inside the stellar
cluster are rather complex, the structure of that gas can
be considered as a network of $N_w$ bubbles, or `cells'
formed around the wind-producing stars.  The mean size of the cell is about
$l_w\simeq lN_w^{-1/3}$. Inside each cell, the wind gas is comparatively cold
($\sim 10^4$ K) and moves away from the star until it collides with the
ambient gas. The latter, being a mixture of
the stellar winds outflowing from the massive stars, is rather hot
($\sim 10^6-10^7$ K, see Sect.4) due to collisions of many individual
stellar winds. Whenewer stellar wind from a massive star collides with the
ambient gas, a shock wave is formed in the wind and the outflowing gas is 
heated.
The  shocked gas joints the hot ambient gas and moves away from the central
cluster through the labyrinth formed by the network of cells. The walls of
this labyrinth are the shock fronts.
Since the wind velocities exceed, by a factor of ten,  the velocity
dispersion of massive stars,
the gas flow in the central cluster can be considered as
quasi-stationary. The structure of the interstellar matter inside the stellar
cluster varies on a time scale of $\sim l_w/\sigma_v\sim 30$ yr.

\bigskip
\noindent{\bf 3.2. Particle acceleration by multiple shocks} 

\noindent {\bf and radiation of the accelerated particles}
\smallskip

Multiple shocks produced by the colliding winds accelerate particles to high
energies until the particles escape from the cluster.
Diffusive acceleration of particles by multiple shocks
produces a distribution in momentum, which is flatter than that
produced by a single shock (e.g., Melrose and Pope 1993); specifically,
the distribution
approaches $f(p)\propto p^{-b}$, with $b=3$ for a large number of
strong shocks, while $b= 4$ for a single strong shock.
A synchrotron spectrum $I_\nu \propto \nu^{-\alpha}$ with
$\alpha =0$ corresponds to the distribution function of relativistic
electrons with $b=3$. The following simplifications have been made
to get this asymptotical distribution of accelerated particles:
(i) all shocks are identical,
(ii) decompression occurs between each shock, and (iii)
at each shock an injection of new particles takes place. In a real situation,
the spectral index may differ from $b=3$. Yet, it is natural to expect
that the synchrotron spectrum of high energy electrons accelerated by
an ensemble of shocks is more or less flat, i.e. $\alpha\simeq 0$.

Synchrotron radiation of high energy electrons produced in the cluster
has to be supressed, owing to the Razin-Tsytovich
effect (Tsytovich 1951; Razin 1960), at low frequencies $\nu < \nu_
{_{\rm RT}}$ such that
$$\nu_{_{\rm RT}}\simeq 20(n_e/B)\,\,\,\,{\rm Hz}\,,\eqno (10)$$
\noindent
where the electron number density, $n_e$, is given in cm$^{-3}$, and $B$
in Gauss. For $n_e\simeq 2\times 10^4$ cm$^{-3}$ and $B\simeq
2\times 10^{-4}$ Gauss this yields $\nu_{_{\rm RT}}\simeq 2$ GHz.

The expected synchrotron luminosity at $\nu > \nu_{_{\rm RT}}$ is given by
$$L_s\simeq \xi {\tau_{\rm esc}\over \tau_s}L_{\rm wind}\,,\eqno (11)$$
\noindent
where $\xi$ is the efficiency of electron acceleration by shocks,
$L_{wind}={1\over 2}\dot Mv^2\simeq 3\times 10^{38}$ ergs s$^{-1}$,
$\tau_{\rm esc}\simeq l/v\simeq 4\times 10^9$ s
is the characteristic time of escape of high energy electrons from
the cluster, and
$$\tau_s\simeq {5\times 10^8\over B^2\gamma_e}\,\,\,{\rm s}\eqno (12)$$
\noindent
is the characteristic time of synchrotron losses
by relativistic electrons of the Lorentz factor $\gamma_e$. Since
 the Lorentz factor of high energy electrons, whose
synchrotron emission in the field $B$ has a maximum
at the frequency $\nu$, is given by
$$\gamma_e\simeq 2\times10^3\left({B\over 10^{-4}\,{\rm Gauss}}
\right)^{-1/2}\left({\nu\over 1\,{\rm GHz}}\right)^{1/2}\,,\eqno (13)$$
one gets from eqs. (11)-(13), for $B\simeq 2\times 10^{-4}$ Gauss, $\xi
\simeq 0.1$, and $\nu\simeq 5$ GHz that $L_s\simeq
3\times 10^{34}$ erg s$^{-1}$, which is comparable to the total radio
luminosity of the compact radio source Sgr A$^\star$. To be more accurate,
the value of $L_s$ should be considered as
an upper limit to the synchrotron emission of high energy electrons,
which are accelerated by an ensemble of shocks in the stellar cluster.

Electrons that are accelerated in the core can serve as a source
of non-thermal X-ray and $\gamma$-ray emissions due to the Compton scattering
of the respective IR and UV photons off the electrons. The
spectrum of those emissions is expected to be flat, like the synchrotron
spectrum discussed above.
The non-thermal luminosity due to the scattered photons is roughly given by
$$L_c\simeq \left({\epsilon_{\rm ph}\over B^2/8\pi}\right)L_s\,,
\eqno (14)$$
where
$$\epsilon_{\rm ph} \simeq {L_{\rm IR+UV}\over 4\pi r_c^2c}\eqno (15)$$
is the photon energy density in IR and UV ranges inside the cluster, and
$L_{\rm IR+UV}$ is the total ${\rm IR + UV}$ luminosity of the cluster.
Certainly,
the value of $L_c$ cannot exceed $\xi L_{\rm wind}\simeq$ a few
$\times 10^{37}$ erg s$^{-1}$. For the same
parameters of the cluster and $L_{\rm IR+UV}\simeq 10^{40}$ erg s$^{-1}$,
eqs. (14) and (15) yield $L_c\simeq 5\times 10^{36}$ erg s$^{-1}$.
Compton scattering of the photons having energies $\sim 0.1-1$ eV
off the electrons having $\gamma_e\sim 10^3$ produces the photons of the mean
energy about $10^2-10^3$ MeV, which is in the {\sl EGRET} range of the
{\sl Compton Gamma Ray Observatory}.
Our estimated value of $L_c$ is comparable to, but somewhat below than, 
the obseved value of $L_\gamma\sim 10^{37}~{\rm erg~s^{-1}}$ (Mattox et 
al. 1993, Mayer-Hasselwander et al. 1993).

Shocks from the colliding winds in the stellar cluster at the Galactic center
can accelerate, besides electrons, protons and nuclei  as well.
The decay of pions that emerge from baryon -- baryon collisions results in
the production of gamma-rays. Before escaping from the stellar cluster,
relativistic nuclei pass through the gas of column density $\Sigma_r\sim
n(l/v)c$.
Putting $n\simeq 2\times 10^4$ cm$^{-3}$, $l\simeq 0.1$ pc, and $v\simeq 700$
km s$^{-1}$, one gets  $\Sigma_r\sim 10^{24}$ cm$^{-2}$.
The cross section of $\pi ^0$-meson production
is about $3\times 10^{-26}$ cm$^2$ for colliding protons near
the threshold of this process, i.e. at $E_p\sim 0.3-1$ GeV, and is larger for
nuclei. Therefore, for $\Sigma_r\sim 10^{24}$ cm$^{-2}$
the efficiency of pion production is somewhat between a few \%
and 10\%. Such an  efficiency seems plausible for conversion of the energy of
accelerated protons and nuclei into the energy of $\gamma$-rays.

Before going out from the stellar core, the outflowing gas has to pass
through one or more shocks. Therefore, one can
expect that the luminosity of the core in cosmic rays might be as high as
$$L_{CR}\sim 0.1~ L_{wind}\sim 3\times 10^{37}~{\rm erg~ s^{-1}}\,.
\eqno (16)$$
The luminosity in $\gamma$-rays owing to
pion decay can reach $L_\pi\sim 0.1~L_{CR}\sim$
a few $\times 10^{36}$ ergs s$^{-1}$. For $\gamma$-rays from pion decays,
there should be a low-energy spectral cutoff at the energy  $E_{\gamma, c}
\sim 70$ MeV. At energies $E\gg E_{\gamma, c}$, the spectral
index of $\gamma$-rays is expected to be similar to the
spectral index of accelerated nuclei. The total luminosity of the
core in $\gamma$-rays might be as high as $\sim 10^{37}$ ergs s$^{-1}$
within a factor of 2 or so. The spectral observations of
$\gamma$-rays from the core could distinguish between the
inverse Compton $\gamma$-ray emission and $\gamma$-rays generated
by pion decays. This would allow to test the above picture and to estimate
the model parameters.
\blankline 
\noindent{\bf 4. DISCUSSION}
\blankline

Colliding winds in the core of the Galactic center's stellar cluster
have two basic differences from those in massive WR + OB binaries. First,
one cannot expect to keep such binaries in the core till the present time
[because in all `relic' massive binaries which might have formed during the
latest starburst in the Galactic nucleus 5-8 Myr ago, at least one component
has already evolved into a compact remnant (Lipunov et al. 1996), whereas
formation and hardening of new hard binaries via triple or double
encounters of massive stars would require a time that exceeds  much the
life-time of those stars on the main sequence]. Nevertheless, a very high 
stellar density in the core makes it feasible for the
encounters between the wind producing stars to be so close as to provide
strong collisions between their winds.
Second, the winds from individual stars in the cluster core propagate
through a rarefied, hot ambient gas that is produced by a
superposition of many individual stellar winds.

Although this environment
looks different from that in massive binaries, it does not influence
the conditions under which, at a close encounter of two stars in the core,
their winds collide so as to produce a strong X-ray emission: Indeed,
the termination shock radius, $R_s$, relative to an individual star
with mass-loss rate $\dot M_w$ and velocity of the wind $V_w$
is given by the condition that the pressure downstream from the shock
equals the sum of thermal and ram pressures of the ambient gas, i.e.,
$$ {\dot M_wV_w\over 4\pi R_s^2}=p_{th} +\rho v^2.  \eqno (17)$$
The central value of the pressure in the stellar core is of thermal 
origin but in the bulk of the core and elsewhere in the central parsec 
the ram pressure dominates the pressure of the outflowing gas. At a 
characteristic radius $r$ from the center of the core, the distance of the 
termination shock from the star is found to be $R_s\approx\left(\dot M_w V_w/ 
\dot M v\right)^{1/2}r$. At $\dot M_w=10^{-6}$\moyr, 
$V_w^\infty=10^3$ km s$^{-1}$, and $\dot M$ and $v$ given by Eq. (1), this  
yields $R_s\approx 0.01$ pc at $r\simeq 0.3$ pc. Thus $R_s$
is much larger than the characteristic
distance of $10^{13}$ cm, at which one expects the X-ray luminosity that
is about the total soft X-ray luminosity of the Galactic center.
In other words, the ambient ram pressure is  
not too large to have any influence on the colliding winds when the impact
parameter is $D\lax 10^{15}$ cm.

At the same time, the ambient ram pressure is large enough to provide a high 
dynamic importance in the entire region occupied by the shocked winds.
As a particular example, let us consider the origin of the wind thought
to be responsible for the dynamics of the Galactic center `mini-cavity'
(for a review of the `mini-cavity'~'s morphology and dynamics, see Genzel
et al. 1994).
Several hypotheses have been explored: (i) winds from IRS 16 complex
are deflected and collimated by the gravitational field of \Sgr
(Wardle \& Yusef-Zadeh 1992), (ii) winds due to one or several stars
very close to the center of the mini-cavity
(Lutz et al. 1993), and (iii) a wind that comes directly from \Sgr
in the form of a jet (Lutz et al. 1993). A recent finding
that \Sgr lies very close to or even within the core of the central
stellar cluster (Eckart et al. 1993) and an enhanced density of stellar
winds in the cluster core both enable us to explore one more possibility:
the wind that provides the energetics of the `mini-cavity' originates
in the cluster core. A presence within the core of even $N_w=10$
wind producing stars with a terminal outflow velocity $V_w^\infty=10^3$
km s$^{-1}$ and an outflow rate $\dot M_w=10^{-5}$\moyr each
(or $N_w=100$ with $\dot M_w=10^{-6}$\moyr) could provide a total wind
mechanical luminosity of about $3\times 10^{37}$ erg s$^{-1}$, which is
consistent
with the estimate that follows from the dynamics of a spherically-symmetric
wind bubble (Lutz et al. 1993). However, an off-center location of the
`mini-cavity' does not easily fit in with a model of a symmetric wind
source centered on the dynamic center. Furthermore, if $N_w$ is too
large, that would  change the present morphology of the `mini-cavity'
and induce a similar cavity elsewhere. Therefore, the  `mini-cavity'
phenomenon could be used for further constraining the number
of wind-producing stars in the core.

Collisions of hot winds at the Galactic center result
in (i) formation of an ensemble of shock waves, (ii) particle acceleration
by the shocks, and (iii) generation of hard X-ray and $\gamma$-ray
emission just as in massive binaries. Although
neither the origin nor the total number of massive, wind producing stars in
the cluster core is known as yet, the production of a strong emission by
the colliding winds (Sec. 3) could, in principle, be used to constrain the
value of $N_w$. However, as is shown in Sec. 2, the rate of wind collisions
able to produce each $L_X\simeq 10^{35}$ erg s$^{-1}$ is only ${\cal R}\sim
10^{-4}$ yr$^{-1}$ if $N_w\sim 10^3$.
Accounting for the influence of a putative massive black hole thought to
present at the Galactic center as a compact radio source \Sgr
would increase this rate by a factor of 30-40 if the black hole mass is
as large as $M_h\sim 10^6$\mo. We also note in passing that the clumps
produced by the colliding winds being accreted onto the black hole would
have induced a strong flaring activity of \Sgr. The recent finding of a
clustering of luminous stars in the immediate vicinity
of \Sgr (Eckart et al. 1995) might indicate an enhanced rate of wind
collisions close to this enigmatic source.

Let us discuss the expected hard radiation resulting from acceleration of
particles by the shock-wave ensemble in the cluster of massive stars at the
Galactic center. As is shown in Sect.~3, acceleration
of particles at shocks produced by the colliding winds at the Galactic
center's stellar core, might result in continuum $\gamma$-ray emission
as high as $L_\gamma\sim 10^{37}$ erg/s. Besides, some $\gamma$-ray lines
could be expected as well. To evaluate the intensity of that emission,
it is helpful to address to the $\gamma$-ray line emission from the Orion
complex recently claimed to be detected
in the 3-7 MeV range (Bloemen et al. 1994). Those $\gamma$-ray lines
have been interpreted as the de-excitation lines $^{12}{\rm C}^*$(4.43
MeV) and $^{16}{\rm O}^*$(6.13 MeV) due to either cosmic-ray carbon and
oxyden nuclei interacting with the ambient medium (Bloemen et al. 1994,
Bykov \& Bloemen 1994) or a combination of those cosmic-ray nuclei plus
cosmic-ray protons and $\alpha$-particles interacting with the ambient
C and O nuclei (Ramaty et al. 1995). Although a recent re-evaluation of
the data indicates that their original interpretation meets serious
problems (Bloemen et al. 1997), it is instructive to estimate the expected
$\gamma$-ray emission associated with the colliding winds at the Galactic
center and then to compare it with the Orion data.
The column density of gas in the stellar core at the Galactic center is
estimated from Eq. (1) to be $\Sigma\simeq 0.6\times 10 ^{22}~{\rm cm}^{-2}$,
which is larger, by a factor of $10^2$, than
that in the Orion complex as given by Cowsik \& Friedlander (1995). Our
upper limit to the number of early-type stars in the vicinity of the
Galactic center's stellar core ($N_w<$100-200, see Sect.2.3) is larger than
(although comparable to) the number of hot massive stars in the Orion
($N=56$, according to Genzel \& Stutzki 1989). Accordingly, the total
kinetic luminosity of the winds from massive stars in the core is
more than that in the Orion complex. (The expected
luminosity in $\gamma$-rays is proportional to the product of $\Sigma L_w$.)
Since the distance to the Galactic
center exceeds that to the Orion complex (about 450 pc) by a factor of 20,
we expect that the flux of $\gamma$-rays from the Galactic center in the
3-7 MeV range is comparable to that from the Orion complex. This conclusion
is consistent with the {\sl SMM} measurements (Harris et al. 1995) that
the total Galactic center (from its central radian) $\gamma$-ray flux
is no greater than 4.7 times the published {\sl COMPTEL} flux from Orion
at 4.4 MeV. Therefore, $\gamma$-rays
from the Galactic center's stellar core might be a target to search
for in further missions such as {\sl INTEGRAL}. Detecting or constraining
that $\gamma$-ray emission, including the line emission
of heavy nuclei such as carbon and oxygen, at 4.4 and 6.1 MeV,
would be important to further constrain the parameters of massive
stars in the cluster at the Galactic center.

Finally, it is worth mentioning that the winds colliding at comparatively
large impact parameters are producing rather strong radio emission.
It has recently been observed that in the vicinity of \Sgr there are
blobs of ionized plasma detected by their radio emission and claimed to be
either ejected from this source (Yusef-Zadeh \etal\ 1990) or resulted from
a not yet specified interaction of the global wind with the gravitational
potential of \sgr (Wardle \& Yusef-Zadeh 1992).
One more possibility that needs to be explored in more detail is that those
blobs could be
the products of the colliding winds in the vicinity of \Sgr. The inferred
parameters of the blobs, $n\simeq  10^{4}~{\rm cm}^{-3}, ~ T\simeq 10^4$ K
(Yusef-Zadeh \etal\ 1990), imply that these blobs are already in the
thermal equilibrium with the shocked wind, $p_{th}\sim 10^8~{\rm
cm^{-3}K}$ (Genzel et al. 1994).

Among those observations which could make further progress in our
understanding of the origin of the wind in Sgr A*/IRS 16 complex, we would
like to mention subarcsecond radio
continuum maps of mass-losing stars in the vicinity of Sgr~A$^*$.
Such observations would help
to resolve some problems related to the origin of the wind, including
the type of wind producing stars. The standard
model for a spherically symmetric wind of partially ionized gas
(molecular weight $\mu$, average ionic charge ${\bar Z}$) with
$T\simeq 10^4$~K and $\dot M\simeq 6\times 10^{-5}$\moyr~
expanding with a constant velocity $v$ would give the flux density of free-free
radio emission
$$\eqalignno{
F_\nu\approx5\,{\rm mJy}\left({\mu\over 1.2}\right)^{-4/3}{\bar Z}^{2/3}
&\left({{\nu}\over{100\,{\rm GHz}}}\right)^{0.6}
\left({{\dot M}\over{6\times 10^{-5}\,M_\odot\,{\rm yr}^{-1}}}\right)^{4/3}
\cr
&\times\left({{v}\over{700\,{\rm km\,s}^{-1}}}\right)^{-4/3}
\left({{d}\over{8.5\,{\rm kpc}}}\right)^{-2}&(18)\cr}$$
(Panagia and Felli 1975; Wright and Barlow 1975).  Measuring
$F_\nu$ at millimeter wavelengths with VLA interferometry
would yield $\dot M$, which with the use of the empirical
relationship $\dot M\propto L^{1.25}$ (van Buren 1985)
would give $L$ and therefore important information about the spectral
types of stars responsible for individual winds.
\blankline 
\noindent{\bf Acknowledgements}
\blankline 
\noindent
We are grateful to an anonymous referee for helpful comments.
{
\par\vfill\eject
\centerline{Table 1}
\centerline{\bf Luminosity and Mass-loss Rate of an Average Wind Producing 
Star}
\blankline
$$\vbox{\offinterlineskip
\halign{&\vrule#&\strut\ #\ \cr
\multispan{7}\hfil\ Table 1\hfil\cr
\noalign{\medskip}
\noalign{\hrule}
height3pt&\omit&&\omit&&\omit&\cr
&\hfil$N_w$\hfil&&\hfil$L/{\rm L}_\odot$\hfil&&\hfil
$\dot M_w/{\rm M_\odot yr^{-1}}$ \hfil&\cr
height3pt&\omit&&\omit&&\omit&\cr
\noalign{\hrule}
height3pt&\omit&&\omit&&\omit&\cr
&\hfil $10^2$\hfil&&\hfil $3\cdot 10^5$\hfil&&\hfil $6\cdot 10^{-5}$
\hfil&\cr

&\hfil $10^3$\hfil&&\hfil $3\cdot 10^4$\hfil&&\hfil $3\cdot 10^{-6}$
\hfil&\cr

&\hfil $10^4$\hfil&&\hfil $3\cdot 10^3$\hfil&&\hfil $2\cdot 10^{-7}$
\hfil&\cr
height3pt&\omit&&\omit&&\omit&\cr
\noalign{\hrule}
\cr}}$$
\par\vfill\eject

\noindent{\bf REFERENCES}
\parskip 0pt
\def\apj    { ApJ{\rm,}\ }

\def\aa     { A\&A{\rm,}\ }

\def\baas   {BAAS{\rm,}\ }
\def\mnras  { MNRAS{\rm,}\ }
\def\ref#1  {\noindent \hangindent=24.0pt \hangafter=1 {#1} \par}
\def\vol#1  { {#1}{\rm,}\ }

\ref{Abbott, D.C. \& Conti, P.C. 1987, ARA\&A 25, 113 }
\ref{Allen, D.A., Hyland, A.R., \& Hillier, D.J. 1990, \mnras\vol
 {244} 706}
\ref{Bloemen, H. 1996 (to be published)}
\ref{Bloemen, H.  et al. 1994, Astron. Astrophys. 281, L 5}
\ref{Bloemen, H.  et al. 1997, ApJ 475, L25}
\ref{Blum, R.D., Sellgren, K., \& DePoy, D.L. 1995, ApJ 440, L 17}
\ref{Bykov, A.  \& Bloemen, H. 1994, Astron. Astrophys. 283, L1   }
\ref{Chlebowski, T. 1989, ApJ 342, 1091}
\ref{Cowsik, R. \& Friedlander, M.W. 1995, ApJ 444, L 29}
\ref{Dougherty, S.M., Williams, P.W., van der Hucht, K.A., Bode, M.F. \&
Davis, R.J. 1996, MNRAS 280, 963}
\ref{Eckart, A., Genzel, R., Hoffmann, R., Sams,
B.J., \& Tacconi-Garman, L.E. 1993, ApJ, 407, L 77}
\ref{Eckart, A., Genzel, R., Hoffmann, R., Sams,
B.J., \& Tacconi-Garman, L.E. 1995, ApJ, 445, L 23}
\ref{Eichler, D. \& Usov, V.V. 1993, ApJ, 402, 27}
\ref{Forrest, W.J., Shure, M.A., Pipher, J.L., \& Woodward, C.A. 1987, in
``The Galactic Center", ed. D.C. Backer. AIP Conf. Proc. 155, 153}
\ref{Friend, D.B. \& MacGregor, K.B. 1984, ApJ 282, 591}
\ref{Geballe, T.R., Persson, S.E., Lacy, J.H., Neugebauer, G., \& Beck,
S.C. 1982, in ``The Galactic Center", ed.\ G.R.~Riegler \&
R.D.~Blandford (New York: Am. Inst. Phys.), p. 60}
\ref{Geballe, T.R., Krisciunas, K.L., Lee, T.J., Gatley, I., Wade, R.,
Duncan, W.D., Garden, R., \& Becklin, E.E. 1984, \apj\vol{284} 118}
\ref{Geballe, T.R., Wade, R., Krisciunas, K.L., Gatley, R., \& Bird, M.C.
1987, \apj\vol{320} 562}
\ref{Geballe, T.R., Krisciunas, K., Bailey, J.A., \& Wade, R. 1991,
ApJ 370, L 73}
\ref{Genzel, R., Hollenbach, D., \& Townes, C.H. 1994, Rep. Progr. Phys.
57, 417}
\ref{Genzel, R. \& Stutzki, J. 1989, ARA\&A 27, 41}
\ref{Genzel, R., Thatte, N., Krabbe, A., Kroker, H., Tacconi-Garman, L.E.
1996, MPE Preprint 362$=$ApJ (submitted)}
\ref{Hall, D.N.B., Kleinmann, S.G., \& Scoville, N.Z. 1982, ApJ 262, L 53}
\ref{Harris, M.J., Share, G.H., \& Messina, D.C. 1995, ApJ 448, 157}
\ref{Krabbe, A., Genzel, R., Drapatz, S., \& Rotaciuc, V. 1991, ApJ 382,
L 19}
\ref{Krabbe, A., Genzel, R., Eckart, A., Naharro, F., Lutz, D., Cameron, M.,
Kroker, H., Tacconi-Garman, L.E., Thatte, N., Weitzel, L., Drapatz, S.,
Geballe, T., Sternberg, A.,\& Kudritzki, R. 1995, ApJ 447, L 95}
\ref{Leitherer, C. 1988, ApJ 326, 356}
\ref{Lipunov, V.M., Ozernoy, L.M., Popov, S.V., Postnov, K.A., \&
Prokhorov, M.E. 1996, ApJ 466, 234}
\ref{Luo, D., McCray, R., \& Mac Low, M.-M. 1990, ApJ 362, 267}
\ref{Lutz, D., Krabbe, A., \& Genzel, R. 1993, ApJ 418, 214}
\ref{Mattox, J.R. et al. 1993, BAAS 24, 1296} 
\ref{Mayer-Hasselwander, H.A. et al. 
1993, Proc. 23rd Intern. Cosmic Ray Conf. 1, 147}
\ref{Melia, F. 1992, ApJ, 387, L25}
\ref{Melrose, D.B. \& Pope, M.H. Proceedings ASA 10, 222, 1993}
\ref{Najarro, F., Hillier, D.J., Kudritzki, R.P., Krabbe, A., Lutz, D.,
Genzel, R., Drapatz, S., \& Geballe, T.R. 1994, A\&A 285, 573}
\ref{Narayan, R., Yi, I., \& Mahadevan, R. 1995, Nat 374, 623}
\ref{Owocki, S.P. \& Rybicki, G.B. 1985, ApJ 299, 265}
\ref{Ozernoy, L.M. 1989, in  ``The Galactic Center", Proc.\ IAU Symp.\
No.~136, ed.\ M.~Morris (Dordrecht: Reidel), p. 555}
\ref{Ozernoy, L.M. 1993, in ``Back to the Galaxy", eds. S. Holt \& F. Verter.
AIP Conf. Proc., 278, 69 }
\ref{Ozernoy, L.M. 1994, in ``Nuclei of Nearby Galaxies: Lessons from
the Galactic Center", eds. R. Genzel and A. Harris (Kluwer Acad. Publ.),
p. 431}
\ref{Ozernoy, L.M. \& Genzel, R. 1996a, in ``The Unsolved Problems of the
Milky Way", Proc. IAU Symp. No. 169, ed. L. Blitz \& P. Teuben 
(Kluwer Acad. Publ.), p. 181 }
\ref{Ozernoy, L.M. \& Genzel, R. 1996b, to be submitted}
\ref{Panagia, N. \& Felli, M. 1975, \aa\vol{39} 1}
\ref{Pollock, A.M. 1987, ApJ 320, 283}
\ref{Pollock, A.M. 1995, in ``Wolf-Rayet Stars: Binaries, Colliding Winds,
Evolution", Proc. IAU Symp. No. 163, eds. K.A. van der Hucht \& P.M.
Williams (Kluwer Acad. Publ.), p. 429}
\ref{Prilutskii, O.F. \& Usov, V.V. 1976, Sov. Astr. 20, 2}
\ref{Ramaty, R., Kozlovsky, B., \& Lingenfelter, R. 1995, ApJ 438, L 21}
\ref{Razin, V.A. 1960, Radiofizika 3, 584}
\ref{Stevens, I.R. 1995, in ``Wolf-Rayet Stars: Binaries, Colliding Winds,
Evolution", Proc. IAU Symp. No. 163, eds. K.A. van der Hucht \& P.M.
Williams (Kluwer Acad. Publ.), p. 486}
\ref{Stevens, I.R., Blondin, J.M., \& Pollock, A.M.T. 1992, ApJ 386, 265}
\ref{Tamblyn, P., Rieke, G.H., Hanson, M.M., Close, L.M., McCarthy, D.W.,
\& Rieke, M.J. 1996, ApJ 456, 206}
\ref{Tsytovich, V.N. 1951, Vest. Mosk. Univ. Phys. 11, 27}
\ref{Usov, V.V. 1990, Ap\&SS 167, 297}
\ref{Usov, V.V. 1992, ApJ 389, 635}
\ref{Usov, V.V. 1995, in ``Wolf-Rayet Stars: Binaries, Colliding Winds,
Evolution", Proc. IAU Symp. No. 163, eds. K.A. van der Hucht \& P.M.
Williams (Kluwer Acad. Publ.), p. 495}
\ref{Usov, V.V. \& Melrose, 1992, D.B. ApJ, 395, 575}
\ref{van Buren, D. 1985, \apj\vol{294} 567}
\ref{Wardle, M. \& Yusef-Zadeh, F. 1992, Nat 357, 308}
\ref{Wright, A.E., \& Barlow, M.J. 1975, in ``The
Galactic Center", Proc.\ IAU Symp.\ No.~136, ed.\ M.~Morris (Dordrecht:
Reidel), p. 443}
\ref{Yusef-Zadeh, F., Morris, M., \& Ekers, R.D. 1990, \baas\vol{22} 865}
\ref{Yusef-Zadeh, F. \& Wardle, M. 1993, \apj \vol{405} 584}
\ref{Zylka, R., Mezger, P.G., Ward-Thompson, D., Duschl, W.J. \& Lesh, H.
1995, AA, 297, 83}
\end